\documentclass[preprint,aps,nofootinbib]{revtex4}
\usepackage{amsfonts,amsmath,amssymb,amsthm}
\usepackage{latexsym}
\usepackage{bbm,bm}
\usepackage{graphicx}

\newcommand{\bra}[1]{\langle #1 \lvert}
\newcommand{\beq}{\begin{equation}}
\newcommand{\eeq}{\end{equation}}
\newcommand{\beqs}{\begin{eqnarray}}
\newcommand{\eeqs}{\end{eqnarray}}

\begin{document}

\title{Comment on ``Path integral action of a particle with the generalized uncertainty principle and correspondence with noncommutativity''}

\author{DaeKil Park$^{1,2}$\footnote{dkpark@kyungnam.ac.kr} and Eylee Jung$^1$,}

\affiliation{$^1$Department of Electronic Engineering, Kyungnam University, Changwon
                 631-701, Korea    \\
             $^2$Department of Physics, Kyungnam University, Changwon
                  631-701, Korea    
                      }

\begin{abstract}
Recently in [Phys. Rev. D $99$ $(2019)$ 104010] the non-relativistic Feynman propagator for harmonic oscillator system is presented when the generalized uncertainty principle is employed. 
In this short comment it is shown that the expression is incorrect. We also derive the correct expression of it. 

\end{abstract}

\maketitle

Recently, S. Gangopadhyay and S. Bhattacharyya\cite{gangop2019} have derived the non-relativistic Feynman propagators for free and harmonic oscillator systems when the generalized 
uncertainty principle (GUP) is employed and have discussed on their correspondence with noncommutativity. The GUP\cite{kempf94} they used can be summarized as a modified Heisenberg algebra
\begin{equation}
\label{modi-heisenberg}
[Q_i, P_j] = i \hbar (\delta_{ij} + \beta \delta_{ij} {\rm P}^2 + 2 \beta P_i P_j],
\end{equation}
where $\beta$ is a GUP parameter, which has a dimension $(\mbox{momentum})^{-2}$. The modified Heisenberg algebra can be readily represented up to first order of $\beta$ as 
$Q_i = q_i, P_i = p_i (1 + \beta {\rm p}^2)$,  where $\{p_i, q_i\}$ satisfies the usual Heisenberg algebra $[q_i, p_j] = i \hbar \delta_{ij}$.  

The authors of Ref.\cite{gangop2019} have considered the harmonic oscillator system, whose Hamiltonian is  
\begin{equation}
\label{oscillator-GUP}
\hat{H} =  \frac{1}{2 m} P^2 + \frac{1}{2} m \omega^2 X^2 = \frac{p^2}{2 m} + \frac{\beta}{m} p^4 + \frac{1}{2} m \omega^2 x^2 + {\cal O} (\beta^2).
\end{equation}
Without any explicit explanation they presented the following Feynman propagator (see Eq. (41) of Ref. \cite{gangop2019}) of this system in a form:
\begin{eqnarray}
\label{prd}
&& \bra{q_f, t_f}q_0, t_0 \rangle                                                              \\     \nonumber
&&= \sqrt{\frac{m \omega}{2 \pi i \hbar \sin \omega T}} 
\left[ 1 + \frac{3 i \beta \hbar m}{T} - 6 \beta m^2 \left(\frac{q_f - q_0}{T} \right)^2 - \frac{3}{4} \beta m \hbar \omega^2 T \cot \omega T + {\cal O} (\beta^2) \right] e^{\frac{i}{\hbar} S_{cl}},
\end{eqnarray}
where $T = t_f - t_0$ and  $S_{cl}$ is a classical action.  The only comment the authors presented is that the $\omega \rightarrow 0$ limit of Eq. (\ref{prd}) is 
\begin{eqnarray}
\label{k-schrodinger-15}
&&K_F [q_f, t_f: q_0, t_0] = \sqrt{\frac{m}{2 \pi i \hbar T}} \left( 1 + \frac{3 i \beta \hbar m}{T} - \frac{6 \beta m^2 (q_0 - q_f)^2}{T^2} + {\cal O} (\beta^2) \right)                \\    \nonumber
&&   \hspace{5.0cm}  \times \exp \left[ \frac{i m}{2 \hbar T} (q_0 - q_f)^2 \left\{ 1 - 2 \beta m^2 \left( \frac{q_0 - q_f}{T} \right)^2  \right\}   \right],
\end{eqnarray}
which is Feynman propagator for free particle. 
Since Eq. (\ref{prd}) is one of the main results of Ref. \cite{gangop2019} and it can be used in other GUP-related issues, it is worthwhile to check the validity of Eq. (\ref{prd}) more carefully. 
Unfortunately, it is incorrect although it approaches to a correct $\omega \rightarrow 0$ limit. As we will show, this should be changed into 
\begin{equation}
\label{k-schrodinger-13}
\bra{q_f, t_f}q_0, t_0 \rangle = \sqrt{\frac{m \omega}{2 \pi i \hbar \sin \omega T}} \left[ 1 + \beta f(q_0, q_f: T) + {\cal O} (\beta^2) \right] e^{\frac{i}{\hbar} (S_0 + \beta S_1)},
\end{equation}
where 
\begin{eqnarray}
\label{k-schrodinger-14}
&&S_0 = \frac{m \omega}{2 \sin \omega T} \left[ (q_0^2 + q_f^2) \cos \omega T - 2 q_0 q_f \right]                                                         \\     \nonumber
&&S_1 = - \frac{m^3 \omega^3}{32 \sin^4 \omega T}  \Bigg[ \left\{ 12 \omega T + 8 \sin 2 \omega T + \sin 4 \omega T \right\} (q_0^4 + q_f^4)               \\     \nonumber
&& \hspace{1.5cm} -4 \left\{ 12 \omega T \cos \omega T + 11 \sin \omega T + 3 \sin 3 \omega T \right\} q_0 q_f (q_0^2 + q_f^2)                                           \\     \nonumber
&& \hspace{4.5cm}   + 12 \left\{ 4 \omega T + 2 \omega T \cos 2\omega T + 5 \sin 2 \omega T \right\} q_0^2 q_f^2    \Bigg]                                               \\     \nonumber
&& f(q_0, q_f: T) = \frac{3 i \hbar m \omega}{8 \sin^2 \omega T}  \left( 2 \omega T + 5 \sin \omega T \cos \omega T + \omega T \cos 2 \omega T \right)                   \\     \nonumber
&& \hspace{2.8cm} - \frac{3 m^2 \omega^2}{8 \sin^3 \omega T} \Bigg[ 2 \omega T \left\{ 3 \cos \omega T (q_0^2 + q_f^2) -  2 (2 + \cos 2 \omega T )q_0 q_f  \right\}                     \\     \nonumber
&& \hspace{5.0cm} + 10 \sin \omega T (q_0^2 + q_f^2 - 2 q_0 q_f \cos \omega T) - 6 \sin^3 \omega T (q_0^2 + q_f^2)       \Bigg].
\end{eqnarray}
Of course, $S_0 + \beta S_1$ is a classical action. It is straightforward to show that the $\omega \rightarrow 0$ limit of Eq. (\ref{k-schrodinger-13}) also goes to $K_F [q_f, t_f: q_0, t_0]$.

In order to show Eq. (\ref{k-schrodinger-13}) explicitly we note that the Feynman propagator $\bra{q_f, t_f} q_0, t_0 \rangle$ can be derived from Schr\"{o}dinger equation as 
\begin{equation}
\label{k-schrodinger-1}
\bra{q_f, t_f} q_0, t_0 \rangle = \sum_{n} \psi_n (q_f) \psi_n^* (q_0) e^{-(i / \hbar) E_n (t_f - t_0)},
\end{equation}
where $\psi_n (q)$ and $E_n$ are $n^{th}$-order eigenfunction and eigenvalue of Schr\"{o}dinger equation.
The Schr\"{o}dinger equation for the harmonic oscillator system is given by 
\begin{equation}
\label{k-schrodinger-2}
\left[ -\frac{\hbar^2}{2 m} \frac{\partial^2}{\partial x^2} + \frac{\beta \hbar^4}{m} \frac{\partial^4}{\partial x^4}  + \frac{1}{2} m \omega^2 x^2 + {\cal O} (\beta^2)\right] \psi_{n} (x) = E_n \psi_n (x).
\end{equation}
If we treat the GUP term $\frac{\beta \hbar^4}{m} \frac{\partial^4}{\partial x^4}$ as small perturbation, one can derive $\psi_n (x)$ and $E_n$  in a form:
\begin{eqnarray}
\label{k-schrodinger-3}
&&\psi_n (x)                                                                                                           \\    \nonumber
&&= \phi_n (x) +                                                                                                       
 (\beta m \hbar \omega)    \bigg[ \frac{(2 n + 3) \sqrt{(n+1) (n + 2)}}{4} \phi_{n + 2} (x) - \frac{(2 n - 1) \sqrt{n (n - 1)}}{4} \phi_{n - 2} (x)   \\  \nonumber
&& \hspace{.5cm}  +  \frac{\sqrt{n (n - 1) (n - 2) (n - 3)}}{16} \phi_{n-4} (x) - \frac{\sqrt{ (n + 1) (n + 2) (n + 3) (n + 4)}}{16} \phi_{n+4} (x)    \bigg] + {\cal O} (\beta^2)    \\    \nonumber
&& E_n = \left(n + \frac{1}{2} \right) \hbar \omega \left[ 1 + \frac{3 (2 n^2 + 2 n + 1)}{2 (2 n + 1)} (\beta m \hbar \omega) \right] + {\cal O} (\beta^2),
\end{eqnarray}
where $n = 0, 1, 2, \cdots$ and 
\begin{equation}
\label{k-schrodinger-4}
\phi_n (x) = \frac{1}{\sqrt{2^n n!}}  \left( \frac{m \omega}{\pi \hbar} \right)^{1/4} H_n \left( \sqrt{\frac{m \omega}{\hbar}} x \right) \exp \left[ - \frac{m \omega}{2 \hbar} x^2 \right].
\end{equation}
In Eq. (\ref{k-schrodinger-4}) $H_n (z)$ is a $n^{th}$-order Hermite polynomial. We assume $\phi_m (z) = 0$ for $m < 0$. 
Inserting Eq. (\ref{k-schrodinger-3}) into Eq. (\ref{k-schrodinger-1}) one can express $\bra{q_f, t_f}q_0, t_0 \rangle$ as 
\begin{equation}
\label{k-schrodinger-5}
\bra{q_f, t_f}q_0, t_0 \rangle = J + (\beta m \hbar \omega) (K_1 + K_2) + {\cal O} (\beta^2)
\end{equation}
where
\begin{eqnarray}
\label{k-schrodinger-6}
&&J = \sum_{n=0}^{\infty} \phi_n (q_f) \phi_n (q_0) \exp \left[ -\frac{i}{\hbar} \left( n + \frac{1}{2} \right) \hbar \omega T \left\{ 1 + \frac{3 (2 n^2 + 2 n + 1)}{2 (2 n + 1)} (\beta m \hbar \omega) \right\} \right]    \\    \nonumber
&&K_1 = \Bigg[\sum_{n=0}^{\infty} \frac{(2 n + 3) \sqrt{(n + 1) (n + 2)}}{4} \left[ \phi_n (q_f) \phi_{n + 2} (q_0) + \phi_n (q_0) \phi_{n+2} (q_f) \right]                              \\    \nonumber
&&\hspace{0.3cm} - \sum_{n=2}^{\infty} \frac{(2 n - 1) \sqrt{n  (n -1)}}{4} \left[ \phi_n (q_f) \phi_{n - 2} (q_0) + \phi_n (q_0) \phi_{n-2} (q_f) \right] \Bigg] \exp \left[-\frac{i}{\hbar} \left(n + \frac{1}{2} \right) \hbar \omega T \right]   \\    \nonumber
&&K_2 = \Bigg[\sum_{n=4}^{\infty} \frac{\sqrt{n(n-1) (n - 2) (n - 3)}}{16} \left[ \phi_n (q_f) \phi_{n - 4} (q_0) + \phi_n (q_0) \phi_{n - 4} (q_f) \right]                              \\    \nonumber
&& \hspace{2.0cm} - \sum_{n=0}^{\infty} \frac{ \sqrt{(n +1) (n + 2) (n + 3) (n + 4)}}{16} \left[ \phi_n (q_f) \phi_{n + 4} (q_0) + \phi_n (q_0) \phi_{n+4} (q_f) \right] \Bigg].                      \\     \nonumber
&&  \hspace{10.0cm} \times \exp \left[-\frac{i}{\hbar} \left(n + \frac{1}{2} \right) \hbar \omega T \right].
\end{eqnarray}
Using the extended Mehler's formula\cite{integral}
\begin{eqnarray}
\label{mehler-1}
&&\sum_{k=0}^{\infty} \frac{t^k}{k!} H_{k+m} (x) H_{k+n} (y) = (1 - 4 t^2)^{-(m + n + 1) / 2} \exp \left[ \frac{4 t x y - 4 t^2 (x^2 + y^2)}{1 - 4 t^2} \right]          \\    \nonumber
&& \hspace{1.5cm}  \times \sum_{k = 0}^{\min (m,n)} 2^{2 k} k! \left(   \begin{array}{c} m  \\  k   \end{array}   \right)    \left(   \begin{array}{c} n  \\  k   \end{array}   \right)   t^k 
H_{m - k} \left( \frac{x - 2 t y}{\sqrt{1 - 4 t^2}} \right)    H_{n - k} \left( \frac{y - 2 t x}{\sqrt{1 - 4 t^2}} \right),
\end{eqnarray}
one can show
\begin{eqnarray}
\label{k-schrodinger-7}
&&J = \sqrt{\frac{m \omega}{\pi \hbar}} e^{   - \frac{i}{2} \omega T}                                  
\left[ \left( 1 - \frac{3 i}{4} (\beta m \hbar \omega^2 T) \right) + \frac{3 i}{2} (\beta m \hbar \omega^2 T) \frac{\partial^2}{\partial \mu^2} + \frac{3}{2} (\beta m \hbar \omega^2 T) \frac{\partial}{\partial \mu} \right] F(\mu) \Bigg|_{\mu = \omega T}
                                                                                                                                                                                                                                           \nonumber    \\
&& K_1 = -\frac{1}{2} \sqrt{\frac{m \omega}{2 \pi i \hbar}} e^{-\frac{3}{2} i \omega T} \sin \omega T \left( 2 i \frac{\partial}{\partial \mu} + 3 \right) G(\mu) \Bigg|_{\mu = \omega T},
\end{eqnarray}
where
\begin{eqnarray}
\label{k-schrodinger-8}
&&F(\mu) = \frac{e^{\frac{i}{2} \mu}}{\sqrt{2 i \sin \mu}} \exp \left[ \frac{i m \omega}{2 \hbar \sin \mu} \left\{ (q_0^2 + q_f^2) \cos \mu - 2 q_0 q_f \right\} \right]                 \\     \nonumber
&& G(\mu) = \frac{\sqrt{2 i} e^{i \mu}}{\sin \mu} \left[ \frac{i m \omega}{\hbar \sin \mu} \left\{ (q_0^2 + q_f^2) \cos \mu - 2 q_0 q_f \right\} + 1 \right] F(\mu).
\end{eqnarray}
Computing Eq. (\ref{k-schrodinger-7}) explicitly, one can show
\begin{equation}
\label{k-schrodinger-9}
J = \sqrt{\frac{m \omega}{2 \pi i \hbar \sin \omega T}} e^{\frac{i}{\hbar} S_0} \tilde{J}     \hspace{1.0cm}
K_1 =  \sqrt{\frac{m \omega}{2 \pi i \hbar \sin \omega T}} e^{\frac{i}{\hbar} S_0} \tilde{K}_1,
\end{equation}
where 
\begin{eqnarray}
\label{k-schrodinger-11}
&&\tilde{J} = 1 - \frac{3 i \beta m \omega^2 T}{8 \hbar \sin^4 \omega T} \Bigg[ -3 i \hbar m \omega (q_0^2 + q_f^2) \sin 2 \omega T + m^2 \omega^2 (q_0^2 + q_f^2 -2 q_0 q_f \cos \omega T)^2      \nonumber    \\
&& \hspace{2.0cm}       + 4 i \hbar m \omega \sin \omega T \left( 2 + \cos 2 \omega T \right) q_0 q_f - \hbar^2 \sin^2 \omega T \left( 2 + \cos 2 \omega T \right)  \Bigg]                          \\     \nonumber
&&\tilde{K}_1 = - \frac{i}{8 \hbar^2 \sin^3 \omega T} \Bigg[ -4 m^2 \omega^2 q_0 q_f (q_0^2 + q_f^2) (3 + \cos  2 \omega T) + 3 \hbar^2 (\cos 3 \omega T - \cos \omega T)                              \\     \nonumber
&&+ 4 m \omega \cos \omega T \left\{ m \omega (q_0^4 + 6 q_0^2 q_f^2 + q_f^4) + 12 i \hbar q_0 q_f \sin \omega T \right\} - 3 i \hbar m \omega (q_0^2 + q_f^2) (5 \sin \omega T + \sin 3 \omega T)   \Bigg].
\end{eqnarray}
Using Eq. (\ref{mehler-1}) again and $H_4 (z) = 16 z^4 - 48 z^2 + 12$, one can show again $K_2 =  \sqrt{\frac{m \omega}{2 \pi i \hbar \sin \omega T}} e^{\frac{i}{\hbar} S_0} \tilde{K}_2$, where
\begin{eqnarray}
\label{k-schrodinger-12}
&&\tilde{K}_2 = - \frac{i \cos \omega T}{16 \hbar^2 \sin^3 \omega T} \Bigg[ 12 m^2 \omega^2 q_0^2 q_f^2 - 3 \hbar^2 (1 - \cos 2 \omega T) + 2 m \omega \bigg\{ m \omega \cos 2 \omega T (q_0^4 + q_f^4)     \\    \nonumber
&&  \hspace{2.0cm}    - 4 m \omega  q_0 q_f (q_0^2 + q_f^2) \cos \omega T
                            - 6 i \hbar \sin \omega T \left\{ (q_0^2 + q_f^2) \cos \omega T - 2 q_0 q_f \right\} \bigg\}    \Bigg].
\end{eqnarray} 
Inserting $J$, $K_1$, and $K_2$ into Eq. (\ref{k-schrodinger-5}), it is possible to show that the Feynman propagator becomes Eq. (\ref{k-schrodinger-13}).


\end{document}